\documentclass[sigconf,screen]{acmart}
\usepackage{cleveref}
\usepackage{microtype}
\usepackage{tabularray}
\UseTblrLibrary{booktabs}

\AtBeginDocument{%
  }

\setcopyright{cc}
\setcctype{by}
\acmDOI{10.1145/3843775.3844549}
\acmYear{2026}
\copyrightyear{2026}
\acmISBN{979-8-4007-2987-4/2026/10}
\acmConference[AISM '26]{Proceedings of the 2nd International Workshop on AI for Software Modernization}{October 12--16, 2026}{Munich, Germany}
\acmBooktitle{Proceedings of the 2nd International Workshop on AI for Software Modernization (AISM '26), October 12--16, 2026, Munich, Germany}
\acmSubmissionID{asews26aismmain-p18-p}
\received{2026-08-11}
\received[accepted]{2026-08-23}

\begin{document}

\title[Recovering Software Architecture Intent from Historical Work Items using Generative AI]{Recovering Software Architecture Intent from Historical Work Items using Generative AI: A Mixed-Methods Industry Case Study}

\author{Dominik Storck}
\orcid{0009-0004-3599-4401}
\affiliation{%
  \institution{Chair of Software Engineering, Technical University of Munich}
  \city{Munich}
  \country{Germany}}
\email{dominik.storck@tum.de}

\author{Tobias Eisenreich}
\orcid{0009-0004-7168-251X}
\affiliation{%
  \institution{Chair of Software Engineering, Technical University of Munich}
  \city{Heilbronn}
  \country{Germany}}
\email{tobias.eisenreich@tum.de}

\author{Stefan Wagner}
\orcid{0000-0002-5256-8429}
\affiliation{%
  \institution{Chair of Software Engineering, Technical University of Munich}
  \city{Heilbronn}
  \country{Germany}}
\email{stefan.wagner@tum.de}

\begin{abstract}
    Software architecture is often only partially captured in code, while much of the design intent lives in evolving project artifacts. In agile projects, work items, user stories, and related tracking documents preserve valuable traces of that intent, but they rarely support direct architectural analysis. This work investigates the recovery of C4 architecture diagrams from historical agile work items using an LLM-based pipeline. The semi-automatic five-step workflow employs a prompt chain, bidirectional traceability, and Chain-of-Thought reasoning to transform unstructured Azure DevOps work items into visual artifacts. Evaluated on two industry projects, we use a mixed-methods design combining qualitative expert interviews with a quantitative stability analysis. Practitioners perceive the generated architectural baselines as accurate and highly useful for system comprehension. Strictly bound by their input data, the artifacts mirror the documented intent, thereby surfacing discrepancies and architectural drift when compared to the implemented reality. Quantitatively, the workflow exhibits high stability for architectural entities but lower stability for their relationships, with relative variance compounding across generation steps. The proposed workflow demonstrates the practical viability of LLM-assisted architectural recovery based on development process artifacts.
\end{abstract}

\ccsdesc[500]{Software and its engineering~Software architectures}
\ccsdesc[500]{Software and its engineering~Software reverse engineering}
\ccsdesc[500]{Computing methodologies~Information extraction}
\ccsdesc[500]{Software and its engineering~Documentation}

\keywords{
Software Architecture Recovery, 
Large Language Models,
Prompt Engineering,
Architectural Erosion, 
Automated Software Documentation, 
Industrial Case Study,
Empirical Software Engineering
}

\maketitle

\section{Introduction}

Software architecture plays a pivotal role in ensuring the quality, maintainability, and scalability of software systems~\cite{Bass2021}. In practice, long-running systems frequently suffer from architectural erosion~\cite{Li2022}. Driven by changing requirements and staff turnover, the implementation can degenerate into an unstructured, convoluted architecture~\cite{Foote1999}. For continued maintenance, it is important to understand the underlying architectural intent.

Traditional Software Architecture Recovery (SAR) approaches attempt to reconstruct system boundaries directly from source code. Yet, empirical studies demonstrate that automated code-level recovery performs poorly, as it captures the eroded reality and implementation workarounds rather than the system's logical design~\cite{Garcia2013}. Conversely, deriving an architecture model from fragmented documentation requires significant manual effort that fundamentally depends on the tacit knowledge of the human architect~\cite{Souza2019}.

To manage architectural complexity, standardized abstraction frameworks like the C4 model~\cite{Brown2026} reduce cognitive load compared to more elaborate modeling languages such as standard UML~\cite{VazquezIngelmo2020}. This makes recovered architectural knowledge more accessible~\cite{Mavrogiorgou2025}, e.g., for modernization planning.

Recent advancements in Large Language Models (LLMs) offer novel deductive approaches to architecture recovery. Yet, empirical observation of the industry reveals a stark barrier to adoption: while organizations actively seek to leverage Artificial Intelligence (AI) for architectural tasks, they continue to struggle with practical implementation due to its non-deterministic nature~\cite{Jahic2024}. By engineering and evaluating an LLM-based workflow that extracts documented architectural intent from historical work items while deliberately bypassing the noise of eroded codebases, we aim to demonstrate the practical viability of Generative AI (GenAI) for architecture rediscovery.

The research presented in this paper was conducted in cooperation with a European technology and consulting company that wishes to remain anonymous. It utilizes real-world project data from the company's internal Azure DevOps environment. As a result, the proposed generation workflow was developed and evaluated under authentic industry conditions. In Azure DevOps, the standard entities representing agile tracking elements, such as Epics, Features, and User Stories, are called work items. In the investigated projects, these work items serve both as agile documentation and as the requirements specification. Therefore, they are our primary proxy for the original system intent. Notably, they capture only what is documented. Hence, we investigate the architectural signal contained in work items, without assuming that this signal is complete.

The LLM-based workflow at the core of this paper is an end-to-end engineering pipeline that translates these historical Azure DevOps work items into structural C4 architecture diagrams. Our research is of exploratory nature and the technical scope is bounded along two axes: the pipeline is driven entirely by textual agile work items, and the generated output is restricted to the upper two tiers of the C4 model: System Context (Level 1) and Container (Level 2) diagrams.
To systematically evaluate the capabilities and limitations of the proposed workflow, we investigate the following research questions:
\begin{itemize}
    \item[\textbf{RQ1:}] How do software practitioners perceive the accuracy and understandability of C4 architecture diagrams automatically generated from fragmented agile tracking data?
    \item[\textbf{RQ2:}] How stable is the workflow's output of C4 System Context and Container diagrams across multiple iterations?
    \item[\textbf{RQ3:}] How can this generative workflow be integrated into modern business processes, and what are its practical utility and limitations?
\end{itemize}

The remainder of the paper is structured as follows: \Cref{section:background} provides the theoretical background, and \Cref{section:related-work} reviews related work. \Cref{section:pipeline} presents the implementation, \Cref{section:methodology} the mixed-methods research design. \Cref{section:results} and \Cref{section:discussion} report and interpret empirical findings, respectively. Finally, \Cref{section:conclusion} summarizes contributions and proposes opportunities for future research.

\section{Background}\label{section:background}
A fundamental challenge in software modernization is architecture erosion, formally defined as the process in which the implemented, as-built architecture increasingly deviates from the intended, as-planned architecture~\cite{Li2022}. Even with optimal maintenance strategies, design erosion is considered inevitable because original design decisions are often lost over time as new requirements necessitate architectural workarounds~\cite{vanGurp2002}. Industry practitioners frequently cite this knowledge loss, alongside missing documentation and developer turnover, as a primary non-technical cause of system degradation~\cite{Li2021Practitioners}.

To comprehend these degraded systems, traditional SAR techniques attempt to extract architectural boundaries from source code using clustering algorithms or dependency analysis. However, code-level recovery techniques achieve surprisingly low accuracy and frequently fail to align with how human architects naturally partition systems~\cite{Garcia2013}. Even when given accurate dependency information, automated tools struggle to produce meaningful architectural representations of large codebases~\cite{Lutellier2015}.

As source code reflects the eroded reality rather than the logical design, standardized architecture abstractions are needed for system comprehension. The C4 model~\cite{Brown2026} advocates for visualizing software at distinct levels of abstraction, allowing architects to communicate designs effectively to different audiences~\cite{Mavrogiorgou2025}. The upper tiers (System Context and Containers) serve as an effective mechanism for defining system boundaries and grouping functional capabilities~\cite{VazquezIngelmo2020}.

\textbf{System Context} -- The System Context Diagram is the most abstract diagram in the C4 model and represents the entire software system embedded into its context.

\textbf{Containers} -- The Container Diagram breaks the system into separately deployable units that execute code or store data, such as databases, microservices, or mobile apps.

With their reasoning capabilities, LLMs are well-suited to process natural language for architectural extraction. However, as task complexity increases, attempting to generate comprehensive and structurally sound outputs in a single pass becomes increasingly prone to data omissions and hallucinations~\cite{schulhoff2025prompt}. One means of mitigation is Chain-of-Thought (CoT) prompting, which generates intermediate reasoning steps before producing a final answer~\cite{wei2023chain}. Prompt chaining links multiple LLM queries sequentially~\cite{Wu2022, schulhoff2025prompt}, breaking the generation process into discrete, manageable steps that reduce the per-step reasoning complexity.

\section{Related Work}\label{section:related-work}
Recent research has begun exploring the application of AI for architecture recovery. Although capable of identifying high-level stylistic patterns, evaluations indicate that state-of-the-art LLMs struggle with precise structural modeling and fine-grained relationship extraction when applied directly to source code~\cite{Amalfitano2026}. Because the codebase itself is often architecturally eroded, direct code-scanning with LLMs inherits the same noise limitations as traditional algorithms.

To bypass this code-level noise, an alternative approach is deductive (top-down) architecture recovery. Rukmono et al.~\cite{Rukmono2024} demonstrated that CoT prompting enables LLMs to reason deductively about architectural properties and classify components using predefined indicators, offering a human-understandable alternative to bottom-up clustering. While their method operates on low-level source code snippets, our workflow applies deductive reasoning directly to historical work items to recover the intended architectural boundaries independently of the eroded implementation.

More broadly, the integration of AI into software architecture primarily focuses on decision support and translating requirements into artifacts~\cite{Esposito2026}. A systematic literature review by Schmid et al.~\cite{Schmid2025} observes that 70\% of existing research relies on basic zero-shot prompting. When attempting advanced sequential prompting, Eisenreich et al.~\cite{EisenreichDDD2026} highlight the risk of error propagation, where minor inaccuracies in the beginning compound over successive steps. Furthermore, Jahić and Sami~\cite{Jahic2024} emphasize that current models lack the determinism required for strict architectural design and frequently produce irreproducible or logically inconsistent results for identical prompts.

Recent studies explore automated diagram generation. Tagliaferro et al.~\cite{Tagliaferro2025} benchmark LLM-generated PlantUML component diagrams against ground truths. Szczepanik and Chudziak~\cite{Szczepanik2025} introduce a multi-agent system for C4 design, evaluated by an LLM judge. However, these works predominantly rely on curated requirement documents or comprehensive system briefs. These pre-structured inputs are designed for experimental purposes and do not reflect the fragmented, inconsistent historical artifacts that practitioners encounter in reality. Although Jahić and Sami~\cite{Jahic2024} explicitly identify non-determinism as a central obstacle, the structural stability of the generated architectures across repeated runs is not systematically reported in these studies.

Prior research lacks a comprehensive evaluation of whether an LLM-based workflow can reliably derive architecture diagrams from historical agile work items. We apply a sequentially chained prompting pipeline to fragmented backlog items from the industry partner's operational project environment. The resulting diagrams are evaluated through both a descriptive stability analysis and qualitative expert feedback along the dimensions of perceived accuracy and understandability (RQ1), output stability (RQ2), and practical integration and utility (RQ3). The primary contributions of this work are twofold: (1) demonstrating that incrementally documented intent can serve as practical baseline for exposing architectural drift without relying on eroded source code, and (2) providing empirical evidence on structural stability and variance propagation in sequential LLM workflows in an industrial context.

\section{Concept and Implementation}\label{section:pipeline}
\begin{figure*}[tbp]
    \centering
     \includegraphics[width=\linewidth]{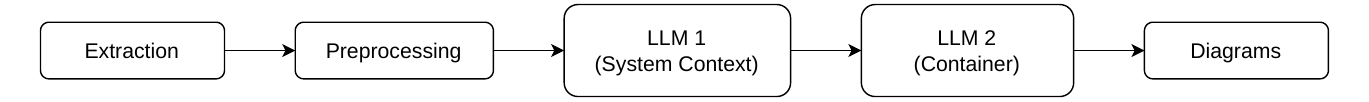}
    \caption{End-to-end execution flow}
    \label{fig:e2e_flow}
\end{figure*}

The developed LLM-based workflow is an end-to-end engineering pipeline that transforms a project backlog into formalized architecture diagrams. As illustrated in Figure~\ref{fig:e2e_flow}, the pipeline operates through five distinct stages, transitioning from raw data extraction to visual rendering. The architectural reasoning is divided into two phases across the two diagrams. This sequential task decomposition aims to reduce the contextual complexity placed on the LLM during each generation step. The structural generation is decoupled from the visual layout by forcing a structured JSON output that can later be converted to the diagramming language.

\textbf{Step 1: Data Extraction.} The pipeline extracts the raw data from Azure DevOps via its REST API, targeting Epics, Features, User Stories, and Tasks to gather sufficient functional context. 

\textbf{Step 2: Preprocessing.} Because raw API responses contain noise, the dataset undergoes strict filtering. To optimize the data for LLM processing, the pipeline converts the API responses into a structured, hierarchical Markdown document, reducing irrelevant content and token overhead.

\textbf{Step 3: LLM 1 (System Context).} A prompt chain~\cite{Wu2022} decouples the architectural generation into two discrete stages. In this step, the LLM processes the Markdown context file and generates the System Context elements (C4 Level 1). The primary objective is to identify the system under design and its boundaries. 

\textbf{Step 4: LLM 2 (Containers).} The validated System Context is injected directly into the Level 2 prompt as additional contextual input in JSON format. This enforces a strict dependency: the LLM must place all newly generated containers strictly within the Stage 1 boundary and reuse exact external element IDs to maintain referential integrity and structural consistency.

\textbf{Step 5: Validation and Visual Rendering.} An intermediate Pydantic validation layer prevents downstream rendering failures. It parses the raw JSON output from both stages to ensure schema adherence. Validated structures are then converted to PlantUML code and transmitted to a local PlantUML server for final visual rendering.

\paragraph{\textbf{Model Configuration \& Prompt Chain}}

Experiments were conducted using a private, EU-hosted Azure OpenAI instance to ensure data privacy for the enterprise dataset. The GPT-5.1 model (version 2025-11-13) was selected during the onset of prototype development in late 2025. To increase determinism, model temperature was set to $0.01$, \texttt{top\_p} remained at its default ($1.0$), and a fixed seed ($42$) was utilized across all API calls~\cite{AzureOpenAICompletions}.

The prompting strategy (Table~\ref{tab:prompt_strategy}) combines schema-guided one-shot prompting, structured CoT, and explicit negative constraints. An injected \texttt{linked\_work\_items} array acts as a bidirectional provenance trail back to the source tickets. The prompts are available in the supplementary repository.

\begin{table}[tbp]
  \caption{Core Prompt Engineering Strategies}
  \label{tab:prompt_strategy}
  \centering
  \begin{tblr}{
    colspec={X[1]X[3]},
    row{1}={font=\bfseries},
    row{2,Z}={halign=l},
    hline{1,Z}={.8pt},
    hline{2}={.5pt},
  }
    {Technique} & {Implementation \& Purpose} \\
    Schema \& Traceability & Injects JSON skeleton with \texttt{linked\_work\_} \texttt{items} array for bidirectional traceability to source tickets. \\
    Structured CoT & Forces \texttt{architecture\_reasoning} as the first key to articulate logic prior to architecture generation. \\
    One-Shot Anchoring & Injects C4 ``Internet Banking''~\cite{Brown2026} example to align output with C4 hierarchy. \\
    Negative Constraints & Uses strict exclusion rules to prevent LLM over-inclusion of non-architectural components. \\
  \end{tblr}
\end{table}

\section{Methodology}\label{section:methodology}

This research follows a mixed-methods design comprising semi-structured expert interviews, a targeted traceability analysis, and a quantitative variance study to measure structural stability.

The study evaluates real-world project data from the industry partner's Azure DevOps environment across two distinct internal projects: Project A (89 work items) and Project B (193 work items). Both projects were in active development for less than one year. The development and calibration of the pipeline used a third, independent project to reduce bias in the evaluation. A fixed baseline set of diagrams was generated before the interviews to ensure that all experts evaluated the same artifacts.

\paragraph{\textbf{Phase 1: Qualitative Evaluation}}
To evaluate the perceived accuracy, understandability, and practical utility of discovered architectures (RQ1, RQ3), we conducted semi-structured interviews with three senior software practitioners from the industry partner: E1 (Project B Tech Lead), E2 (Project A Solution Designer), and E3 (Project A Originator). The sessions followed an interview guide~\cite{Hove2005} available in the supplementary data and used the fixed baseline diagrams as visual elicitation artifacts to ground the discussion. The interview transcripts were analyzed via thematic synthesis~\cite{Cruzes2011}. During the interviews, experts identified specific architectural discrepancies, noting elements that appeared either unexpectedly present or notably absent from the diagrams. Therefore, we additionally performed a targeted traceability analysis to cross-reference these flagged discrepancies with the complete input corpus, thereby verifying the root cause of each divergence.

\paragraph{\textbf{Phase 2: Quantitative Evaluation}}
To assess workflow determinism and stability (RQ2), the pipeline was executed $n=10$ independent times per project under strictly identical configuration parameters (GPT-5.1, temperature 0.01, fixed seed), yielding 40 diagrams in total. Output stability here refers to the extent to which repeated generation from the same input produces structurally similar diagrams.

Following established architecture measurement principles~\cite{Coulin2019}, stability was operationalized through two core structural metrics: node count (entities) and edge count (relationships). For Container diagrams, counting was strictly constrained to internal elements within the \texttt{System\_Boundary} to isolate the LLM's net-new generation at the container level from the inherited context-level elements. Outputs are described using minimum, maximum, mean ($\mu$), median, and mode. While the sample mode is often uninformative or misleading for small samples of continuous data, its application is supported for numeric-discrete variables that span a narrow range of possible values~\cite{Letkowski2013}. In such discrete distributions, the mode serves as a valid measure of location, identifying the point of maximum probability mass~\cite{Dutta2010}. Tracking the mode is therefore analytically valuable for this study. It uncovers the specific architectural configuration the LLM is most prone to generate under fixed conditions. Absolute dispersion is quantified via Standard Deviation (SD / $\sigma$), and relative variance across projects of differing sizes is normalized using the Coefficient of Variation ($CV = \frac{\sigma}{\mu}$). Manual inspection of the generated diagrams across iterations confirmed high semantic overlap of the architectural elements. Therefore, we use variance in absolute node and edge counts as practical measure of the model's structural baseline stability.

\section{Results}\label{section:results}
This section presents the empirical findings, combining the qualitative thematic synthesis of expert interviews with the quantitative structural stability analysis.

\subsection{Qualitative Expert Evaluation}
The thematic synthesis of the expert interviews yielded four final themes, uniformly supported across the expert panel (E1, E2, E3). Table~\ref{tab:summary_of_findings} summarizes these themes. The term \emph{documented intent} refers to the directional content of work items---stated plans whose actual realization remains conditional on subsequent execution and vulnerable to architectural drift.

\begin{table*}[tbp]
  \caption{Summary of findings.}
  \label{tab:summary_of_findings}
  \centering
    \begin{tblr}{
        colspec={X[4]X[8]X[.5]},
        row{1}={font=\bfseries},
        row{2,Z}={halign=l},
        hline{1,Z}={.8pt},
        hline{2}={.5pt},
    }
    Theme & Core Finding & RQ \\
    T1: Resilient but bounded extraction & Output is perceived as accurate despite sparse input, remains strictly bounded by documentation content, and in edge cases surfaces divergences between documented intent and implemented reality. & RQ1 \\
    T2: High perceived utility with purpose-dependent reliability & Substantial utility and quantified time savings for baseline discovery are reported, with human-in-the-loop refinement required for external-facing purposes. & RQ3 \\
    T3: Understandability friction as a rendering, not a semantic, constraint & All observed cognitive friction is attributable to the PlantUML rendering layer; the LLM's semantic output is consistently well understood. & RQ1 \\
    T4: Workflow integration: continuous consistency and architectural auditing & The workflow delivers continuous documentation freshness, a starting canvas for modernization or design milestones, and an auditing function that exposes architectural drift and disciplines ticket hygiene. & RQ3 \\
  \end{tblr}
\end{table*}

\paragraph{\textbf{Theme 1: Resilient but Bounded Extraction.}} 
Practitioners evaluate the recovered diagrams as highly accurate, even when underlying historical work items are fragmented or vague: \textit{``I know our stories and how they're all written a bit vaguely, and that this comes out of that is already pretty good''} (E1). However, the extraction serves as a strict generative mirror. The structural output remains bounded by the input, inheriting any outdated content or organizational quirks of the project management. A direct consequence of this strict bounding property is the tool's diagnostic affordance: by faithfully rendering the documented, as-planned intent, the pipeline visually surfaces where historical requirements diverge from the implemented, as-built reality. Rather than merely reflecting a broken state, experts noted that this diagnostic visibility forces architectural reflection. However, utilizing this auditing mechanism to detect architectural drift requires sufficient tacit project knowledge.

\paragraph{\textbf{Theme 2: High Perceived Utility with Purpose-Dependent Reliability.}}
The diagrams provide substantial utility and quantifiable time savings, reducing the manual creation of architectural baselines from hours to minutes. Yet, adoption is governed by a strict trust boundary: while the raw output is highly effective for internal engineering and rapid knowledge transfer, external or client-facing scenarios mandate manual validation and curation. The recovered diagram acts as an initial baseline rather than a final source.

\paragraph{\textbf{Theme 3: Understandability Friction as a Rendering Constraint.}}
Practitioners found the LLM's semantic extraction (architectural logic, component descriptions, relationships) accurate and consistently well comprehensible. Reported cognitive friction (e.g., inadequate visual differentiation, suboptimal spatial layouts) concerned the PlantUML rendering layer, not the model's semantic reasoning.

\paragraph{\textbf{Theme 4: Continuous Consistency and Architectural Auditing.}} 
The workflow offers three integration dimensions: providing instant, continuous documentation; serving as a starting canvas for project milestones (e.g., design planning or modernization handovers); and functioning as an audit tool. Visually confronting the team with missing or misrepresented elements exposes underlying documentation debt and structural drift. Thereby, the pipeline acts as an incentive mechanism to discipline ticket hygiene for ongoing application understanding and accurate representation. Because manually drawn artifacts often suffer from immediate documentation lag, the experts identified this continuous documentation freshness as the tool's primary value.

\paragraph{\textbf{Analysis of Architectural Discrepancies.}}
Tracing flagged discrepancies back to the Azure DevOps context files (Table~\ref{tab:discrepancies}) confirms the LLM's strict bounding behavior.

The analysis underlines that the LLM successfully extracted \textit{unexecuted intent}, while remaining completely blind to \textit{undocumented implementations}. In a modernization context, this provenance tracking successfully isolates abandoned legacy plans from the actual system state. Furthermore, undocumented implementations (e.g., Cloud Storage in Project A) represent practical manifestations of architectural erosion---code-level additions made outside the formal design-process documentation. Although the \texttt{linked\_work\_items} reference tags offer a reliable traceability chain, some provide only loose semantic backing, making them strong but not infallible.

\begin{table*}[tbp]
  \caption{Traceability analysis of identified architectural discrepancies.}
  \label{tab:discrepancies}
  \begin{tblr}{
    colspec={XX[.5]X[.8]XXX},
    row{1}={font=\bfseries},
    row{2,Z}={halign=l},
    hline{1,Z}={.8pt},
    hline{2}={.5pt},
  }
    Element (Project) & Diagram & Implementation & Expert Suspicion & Context Provenance & Classification \\
    Auth Gateway (A) & Present & Absent & Anticipated feature & Explicitly documented & Unexecuted intent \\
    Cloud Storage (A) & Absent & Present & Undocumented feature & Undocumented & Undocumented implementation \\
    Cloud Storage (B) & Present & Absent & Suggested best practice & Explicitly documented & Unexecuted intent \\
    Email Service (B) & Present & Absent & Abandoned planned feature & Explicitly documented & Unexecuted intent
  \end{tblr}
\end{table*}

\subsection{Quantitative Structural Stability}
\begin{figure}[tbp]
  \centering
  \includegraphics[width=\linewidth]{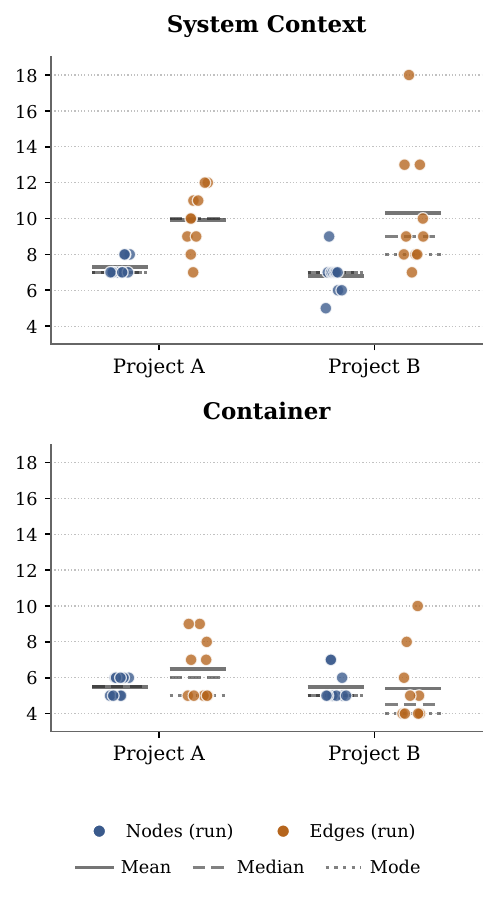}
  \caption{Per-run distribution of node and edge counts across Context and Container phases.}
  \label{fig:c4-stability-stripplot}
\end{figure}

This section quantifies the inherent non-determinism of the workflow's underlying LLM by measuring structural variance across 40 diagrams. All generations utilized a fixed configuration (GPT-5.1, temperature 0.01, fixed seed) to isolate stochastic behavior. Figure~\ref{fig:c4-stability-stripplot} visualizes the per-run distributions.

\paragraph{\textbf{Stability of System Context Generation.}} 
Across both projects, node generation exhibited lower variance than edge generation. Project A node generation was most stable (CV = 6.6\%), anchoring at 7 nodes in 70\% of iterations. Project B showed slightly more variance (CV = 15.2\%) but established a dominant baseline of 7 nodes in 60\% of runs. In contrast, edge generation displayed a broader distribution. Project B edges produced the highest variance of this phase (CV = 33.0\%, ranging from 7 to 18 edges). Project A edges (CV = 16.8\%) were multimodal and likewise failed to converge on any stable configuration. Overall, the model reproduces architectural entities far more reliably than the relationships between them.

\paragraph{\textbf{Stability of Container Generation.}}
The second LLM step mirrors the stability patterns observed in the Context phase. Node generation remained markedly more stable than edge generation. Project A demonstrated a perfectly bimodal split of 5 or 6 nodes (CV = 9.6\%). Project B reproduced a baseline of 5 internal containers in 70\% of iterations (CV = 15.5\%). Internal edges showed higher relative variance, peaking at a CV of 38.3\% for Project B, alongside significant dispersion for Project A (CV = 26.4\%). This indicates that the output quantity for internal container nodes clusters closely around a central value, whereas the corresponding edge counts again exhibit a broader numerical spread.

\paragraph{\textbf{Overarching Structural Patterns.}}
Four distinct structural patterns emerged across both generation steps. 
First, Project A (89 work items) consistently exhibited lower relative variance than Project B (193 work items), indicating that stability baseline levels may be influenced by project-specific input characteristics such as requirement volume, granularity, or domain content. 
Second, absolute variance for node generation demonstrates high practical stability. In three out of the four node configurations, the standard deviation remained below 1.0, with two hovering near 0.5, one at 0.85, and the fourth at 1.03. This indicates that the model typically deviates by less than a single architectural element from its mean node output.
Third, the edge data hints at a right-skewed distribution. For both projects in the Container step, the minimum edge count exactly matches the mode. In three of the four edge configurations, the metrics follow an order in which the mode is less than the median, which in turn is less than the mean (e.g., 4 < 4.50 < 5.40 for Project B Container phase). This skew indicates that the variance is primarily additive; the LLM establishes a frequent baseline of core connections while the variance is driven by iterations generating higher edge counts, which pull the mean upward (visible in Figure~\ref{fig:c4-stability-stripplot} as the staggered Mode--Median--Mean lines for the edge variables).
Finally, the data reveal a consistent increase in relative instability between the two LLM steps: relative variance compounding across the chained generation steps results in a higher CV in the Container phase than in the Context phase.

Together, these findings characterize the workflow's behavior as descriptively stable around dominant baselines, with relative variance concentrated in edge generation and amplified by sequential generation steps.

\section{Discussion}\label{section:discussion}
Synthesizing the empirical results provides comprehensive insights for the evaluation of the proposed LLM workflow.

\paragraph{\textbf{Accuracy and Understandability (RQ1).}}
The qualitative findings indicate that practitioners perceive the automatically generated C4 diagrams as highly accurate reflections of the corresponding software projects and their historical work items. The LLM successfully extracts architectural elements despite natural-language ambiguities and sparse input. The understandability of the generated C4 diagrams is limited by the rendering technology rather than the model's semantic reasoning.

The LLM does not generate an idealized technical architecture but acts as a mirror. It extracts and models precisely what is documented, including the backlog's inherent structural biases. Similar to Conway's Law~\cite{Conway1968}, which states that organizations design systems that reflect their own communication structures, the LLM-driven architectural extraction directly translates the backlog's administrative structure into the architecture's structural outcome. For example, the generated Project A Container diagram misrepresented deployment units by splitting them along team competency lines (e.g., separating backend functionality by developer role). Because the underlying tasks were organized by human resource allocation rather than technical architecture, the LLM mirrored this organizational quirk as a structural boundary. As the traceability analysis underlines, the artifacts reflect the documented intent, even when that intent diverges from the technical reality.

By acting as a strict mirror of that historical intent, the workflow provides a mechanism to detect architecture erosion. Because the LLM derives the architecture solely from work items, it effectively isolates the \textit{as-planned} intent. When practitioners compare this recovered intent with the actual codebase, they actively identify areas of architectural drift. However, utilizing this diagnostic affordance requires the practitioner to actively compare the artifact against their mental model of the software. Thus, the tacit domain knowledge of the human architect remains a strong dependency, aligning with the findings of Souza et al.~\cite{Souza2019}. The workflow does not replace the architect's expertise; rather, it shifts their cognitive effort away from manual diagramming and toward validating exposed structural discrepancies.

\paragraph{\textbf{Workflow Stability (RQ2).}}
The workflow achieves high structural stability, revealing a strong tendency toward a consistent, dominant baseline of nodes. Because generation is automated, users can easily regenerate unsatisfactory artifacts with a high probability of returning to this distinct structural baseline.

Beyond this baseline stability, two specific patterns emerged: the model identifies architectural entities (nodes) much more consistently than their relationships (edges), and stability degrades as the pipeline progresses. This aligns with the observation of Jahić and Sami~\cite{Jahic2024}, who emphasize that LLMs lack the strict determinism required for rigorous architectural design. The quantitative results of this paper specify this limitation: the model is highly consistent regarding \textit{what} exists in the architecture, but remains structurally less consistent regarding exactly \textit{how} those elements interact. This disparity logically aligns with the inherent complexity of mapping software interactions. For instance, across iterations the LLM oscillates between summarizing a reciprocal data flow (e.g., an HTTP request and its response) as a single edge or breaking it down into multiple granular connections. The system prompts do not enforce a strict diagrammatic convention regarding directionality though. Furthermore, the workflow's sequential execution compounds instability. This observation is consistent with error propagation in multi-step LLM workflows~\cite{EisenreichDDD2026}. Minor variations generated in the Context step are passed down as ground truth, causing the second model to operate on a slightly different foundational context in each run, amplifying overall structural inconsistency.

\paragraph{\textbf{Utility, Integration, and Limits (RQ3)}}
The generated artifacts offer substantial utility and can be effectively integrated into modern business processes. The automated pipeline significantly reduces the manual effort traditionally required to compile architectural overviews, cutting baseline creation from hours to minutes. In any effort that requires system understanding, this extraction provides rapid baseline discovery from ticketing history. Beyond creation speed, experts proposed integrating the artifacts as a starting canvas for critical milestones, such as technical kickoffs, infrastructure readiness checks, and project onboarding.

However, deployment is governed by a strict trust boundary. The raw, unedited output of the workflow is highly effective for internal engineering purposes and initial architectural exploration, but insufficient for external, pre-sales, or client-facing use. For example, internally, the raw diagram serves as an immediate catalyst for discussions and allows rapid knowledge transfer. Conversely, before a generated diagram can cross the trust boundary to an external stakeholder, it must undergo manual curation to resolve rendering friction, fill any gaps, and validate the semantic mapping. This human-in-the-loop condition aligns with current literature in which 85\% of studies use GenAI in software architecture as an assistive tool~\cite{Esposito2026}. Furthermore, Abbasi et al.~\cite{Abbasi2025} emphasize that AI must be treated as an augmentative tool whose outputs are rigorously overseen by human experts.

By faithfully rendering incomplete or vague work items into visual artifacts, the workflow also exposes underlying documentation debt. Confronting the development team with tangible consequences of their documentation practices creates a continuous feedback loop that encourages better ticketing discipline. For legacy systems, it simply represents the legacy state. Consequently, the recovered artifacts serve a dual purpose: in cases of divergence, they actively surface architectural drift, while in cases of alignment, they provide an accurate system representation necessary for deep application understanding.

Ultimately, the workflow counters knowledge loss~\cite{vanGurp2002}. By formalizing unstructured agile tracking data into standardized C4 models on demand, the pipeline captures and preserves architecture decisions. Practitioners explicitly noted that manually drawn artifacts are frequently obsolete shortly after active development begins and the first changes are applied. By eliminating the friction of manual diagram maintenance, the workflow operationalizes lightweight architecture modeling~\cite{Jongeling2025}, ensuring that structural documentation realistically keeps pace with the high velocity of modern development. With that, the workflow provides continuous consistency: the ability to instantly reconstruct documented architectural intent from historical records. This enables teams to ground their modernization efforts in the system's documented design logic without being misled by a potentially noisy and eroded codebase.

\paragraph{\textbf{Limitations.}}
Several threats to validity constrain this work. Internally, the expert evaluation introduced familiarity bias, as practitioners were intimately familiar with the evaluated projects. As architecture recovery typically lacks a certified ground truth, validation relied on the experts' mental models, which are susceptible to recall bias. Uninitiated stakeholders might perceive understandability differently. To mitigate recall decay and anchor the discussion, we utilized the generated baseline diagrams as visual elicitation artifacts during the interviews~\cite{Hove2005}. Additionally, qualitative coding was performed by a single researcher, introducing potential interpretive bias. We sought to minimize this by adhering strictly to a defined thematic synthesis protocol~\cite{Cruzes2011}, supplemented by AI-assisted analysis of the raw interview transcripts.

Construct validity represents the primary limitation of our quantitative analysis. We assessed structural stability using absolute node and edge counts. While this proxy captures topological variance and volume, it does not evaluate the semantic correctness or identity of the fluctuating elements. A manual inspection confirmed semantic overlap, but does not quantitatively assess it. Furthermore, we addressed this quantitative limitation through the qualitative expert interviews, which serve to evaluate the semantic accuracy of the generated architectures.

Finally, due to the inherent nature of a single-case study design, external validity is limited. The evaluation relies on three practitioners from a single company and on a single LLM (GPT-5.1). The fidelity of the generated artifacts is inherently bounded by the organization's specific work-item maintenance culture. Additionally, both evaluated projects were relatively young ($\leq 1$ year in development). While we tried to mitigate bias by implementing the pipeline without testing it on the evaluation projects, this work does not support generalizing the findings beyond our case study. Furthermore, scaling this workflow to massive enterprise systems with thousands of tickets is fundamentally limited by LLM context windows, which degrade when processing extensive inputs~\cite{naveed2025}.

\section{Conclusion}\label{section:conclusion}
This paper investigated the semi-automatic recovery of C4 architecture diagrams from historical agile work items, aiming to overcome the limitations of code-level recovery. By evaluating a deductive, five-step generative workflow that extracts system context and container boundaries, this work assessed the structural accuracy, statistical stability, and practical utility of using LLMs to extract architectural intent from project tracking data.

The empirical evaluation demonstrates that the pipeline accurately mirrors the input data, directly reflecting the documented as-planned state of the project. The model exhibits a high tolerance for natural-language ambiguity and successfully extracts useful architectural baselines. Bypassing implementation details, the workflow reliably renders unexecuted intent while remaining blind to undocumented implementations. This provides a diagnostic affordance, allowing practitioners to isolate architectural drift and expose areas of system erosion. The extraction proved practically stable, though entities were identified much more consistently than their relationships. Crucially, sequential prompting compounded structural variance, with minor deviations in the initial context extraction propagating into the subsequent container generation phase. Despite these constraints, the workflow delivers substantial utility for software practitioners. It offers rapid baseline discovery and implicitly mitigates the loss of historical design decisions. However, its deployment is subject to a clear trust boundary: while the raw outputs are highly effective for internal matters, external or client-facing applications require human curation. By continuously recovering documented architectural intent from historical records, it provides software practitioners with an uneroded starting canvas.

\paragraph{\textbf{Future Work}}
The identified limitations present concrete opportunities for future research:
(1) \textit{Formalized Evaluation Frameworks:} Establishing standardized frameworks~\cite{Esposito2026} to objectively assess structural and semantic accuracy and establish comparability within the field.
(2) \textit{Scaling to Enterprise Contexts:} Investigating RAG approaches or hierarchical extraction models to parse thousands of agile tickets without exceeding LLM context windows.
(3) \textit{Multi-Agent Self-Correction:} Transitioning to multi-agent architectures~\cite{Eisenreich2024} with generator-evaluator loops to mitigate the error propagation~\cite{EisenreichDDD2026} observed in sequential prompt chains.

\section*{Data Availability}\label{data_availability}
To comply with corporate confidentiality constraints, the proprietary dataset of Azure DevOps work items used in this study and the generated output diagrams cannot be shared publicly. The source code of the workflow, the iterative prompt history, and the interview guide are available in the supplementary data~\cite{dmnksto_2026_22145614}.

\begin{acks}
The authors used generative AI to assist with grammar, translation of quotes, data visualization, qualitative analysis, code generation, and peer-review simulation. All technical ideas, analyses, results, and conclusions were conceived, developed, and verified solely by the authors, who take full responsibility for the final manuscript.
\end{acks}

\bibliographystyle{ACM-Reference-Format}
\bibliography{references}

\end{document}